\documentclass[preprint,superscriptaddress,amsmath,amssymb,prd,aps,showpacs,floatfix,nofootinbib]{revtex4-1}
\usepackage{graphicx}
\usepackage{dcolumn}
\usepackage{bm}
\usepackage{mathrsfs}
\usepackage{hyperref}
\usepackage{amsmath}
\usepackage{amssymb}
\usepackage{bm}
\usepackage{color}
\usepackage{float}
\usepackage{dcolumn}
\usepackage{multirow}
\usepackage{changepage}
\usepackage{enumerate}
\usepackage{ulem}
\usepackage[dvipsnames]{xcolor}
\usepackage{braket}
\usepackage{nicefrac}
\usepackage{multirow}
\usepackage{tabularx}
\usepackage{subfigure}
\usepackage{soul}
\usepackage[makeroom]{cancel}

\hypersetup{pdftex,colorlinks=true,linkcolor=blue,citecolor=ForestGreen,menucolor=black,urlcolor=blue,filecolor=blue}

\newcommand{\mev}{\textrm{ MeV}}

\begin{document}

\title{$T_{bc\bar s}$ states in the process $\Upsilon \to D^{-} \bar B^{0} D_s^{+}$}
\date{\today}

\author{Wen-Ying Liu}
\affiliation{School of Physics, Southeast University, Nanjing 210094, China}

\author{Hua-Xing Chen}
\email{hxchen@seu.edu.cn}
\affiliation{School of Physics, Southeast University, Nanjing 210094, China}

\begin{abstract}

We perform a theoretical study of the decay process $\Upsilon \to D^{-} \bar{B}^{0} D_s^{+}$ in search of the doubly heavy tetraquark states $T_{bc\bar{s}}$ with quark content $bc\bar{s}\bar{d}$. These $T_{bc\bar{s}}$ states are assumed to be dynamically generated molecular states from the S-wave interactions between $\bar{B}_s^{(*)0} D^{(*)+}$ and $\bar{B}^{(*)0} D_s^{(*)+}$ meson pairs. Based on the total angular momentum and the type of the constituent mesons (pseudoscalars $P$ or vectors $V$), they are labeled as $T_{bc\bar{s}}^{0, PP}$, $T_{bc\bar{s}}^{0, VV}$, and $T_{bc\bar{s}}^{2, VV}$, respectively. The $\bar{B}^{0} D_s^{+}$ invariant mass distribution for this decay is calculated using current algebra, incorporating contributions from $T_{bc\bar{s}}$ states arising from final-state interactions. Our results reveal a clear peak structure in the $7415 - 7425\,\mev$ region, which is attributed to the $T_{bc\bar{s}}^{2, VV}$ state. Additionally, a distinct dip structure appears near the $\bar{B}^{*0} D_s^{*+}$ threshold, characteristic of the $T_{bc\bar{s}}^{2, VV}$ as a hadronic molecular state. A near-threshold enhancement associated with the $T_{bc\bar{s}}^{0, PP}$ state and a dip arising from the $T_{bc\bar{s}}^{0,\,VV}$ state are also identified, though the manifestation of these features depends sensitively on model parameter fine-tuning. Therefore, with increased experimental statistics, the decay channel $\Upsilon \to D^{-} \bar{B}^{0} D_s^{+}$ offers a promising avenue for discovering and characterizing the $T_{bc\bar{s}}$ states.
\end{abstract}

\maketitle

\section{Introduction}
\label{sec:intro}

Quantum chromodynamics (QCD), the fundamental theory of the strong interaction, predicts the existence of conventional hadrons—mesons composed of quark-antiquark pairs and baryons made of three quarks. However, the discovery of candidates of exotic hadrons  \cite{BES:1999dmf,BES:2006nqh,BESIII:2013qqz,BESIII:2020nme,Etkin:1985se,BES:2003aic,BESIII:2010vwa,IHEP-Brussels-LosAlamos-AnnecyLAPP:1988iqi,E852:1998mbq,E852:2004gpn,BESIII:2022iwi,BESIII:2022riz,BESIII:2019wkp}, such as tetraquarks \cite{BESIII:2022bse,Belle:2003nnu,Belle:2004lle,CDF:2009jgo,CDF:2011pep,Belle:2014nuw,CLEO:2006tct,BaBar:2005hhc,Belle:2005lik,BESIII:2013ris,Belle:2014vzn,Belle:2013urd,LHCb:2020pxc,LHCb:2020bls,LHCb:2020bwg,ATLAS:2023bft,BESIII:2020qkh} and pentaquarks \cite{LHCb:2015yax,LHCb:2019kea,LHCb:2021chn,LHCb:2020jpq}, has challenged this traditional picture, revealing a richer structure in the non-perturbative regime of QCD. These states, which cannot be accommodated within the naive quark model, have become central to modern hadron physics, offering unique insights into quark-gluon dynamics and the mechanisms of hadronic spectroscopy. Among them, tetraquark candidates, composed of two quarks and two antiquarks, are of particular interest due to the role of hadronic molecular configurations. The study of such states is essential for completing the QCD hadron spectrum and understanding the dynamics of hadronic formation.

Recent experimental observations, notably the observation of the $T_{cc}$ tetraquark state \cite{LHCb:2021vvq,LHCb:2021auc}, have provided compelling evidence for the existence of doubly heavy tetraquarks. Theoretical interpretations of $T_{cc}$ state span a range of possibilities, including compact tetraquark structures and loosely bound hadronic molecules formed via meson exchange \cite{Yang:2009zzp,Li:2012ss,Ohkoda:2012hv,Guo:2021yws}. Guided by heavy quark symmetry, which suggests universal behavior across systems with different heavy quark content, the existence of  $T_{bc\bar{s}}$ tetraquark states composed of $b$, $c$, $\bar{s}$, and $\bar{q}$ ($q = u,d$) has been proposed. In our previous work \cite{Liu:2023hrz}, we studied the $T_{bc\bar{s}}$ states as hadronic molecules within the extended local hidden gauge approach by solving the Bethe-Salpeter equation for the $\bar{B}_s D$, $\bar{B} D_s$, $\bar{B}_s^* D$, $\bar{B}^* D_s$, $\bar{B}_s D^*$, $\bar{B} D_s^*$, $\bar{B}_s^* D^*$, and $\bar{B}^* D_s^*$ systems. Six dynamically generated poles were found below respective thresholds, with quantum numbers $J^P = 0^+$, $1^+$, and $2^+$, and binding energies of approximately 10–20$\mev$ using a cutoff momentum of $q_\text{max} = 600 \, \mev$. In the present work we adopt the notation $T_{bc\bar{s}}^{J, ab}$, where $J$ denotes the total angular momentum and $a, b$ represent the types of the constituent mesons ($P$ for pseudoscalar meson, $V$ for vector meson). Accordingly, those poles corresponding to the $T_{bc\bar{s}}$ states are labeled as $T_{bc\bar{s}}^{0, PP}$, $T_{bc\bar{s}}^{1, VP}$, $T_{bc\bar{s}}^{1, VP^\prime}$, $T_{bc\bar{s}}^{0, VV}$, $T_{bc\bar{s}}^{1, VV}$, and $T_{bc\bar{s}}^{2, VV}$. Specifically, the $VP$ constituent refers to the $\bar{B}_s^* D$ and $\bar{B}^* D_s$ systems, whereas the $VP^\prime$ constituent corresponds to the $\bar{B}_s D^*$ and $\bar{B} D_s^*$ systems. The study of $T_{bc\bar{s}}$ states is not only crucial for verifying the universality of tetraquark configurations but also for establishing a comprehensive framework for exotic hadron spectroscopy.

Weak decays of heavy hadrons have proven to be fertile grounds for studying exotic hadrons \cite{Oset:2016lyh,Chen:2021erj,Chen:2015sxa,BaBar:2019hzd,BESIII:2024mbf,LHCb:2023eig}. The $\Upsilon$ meson, a bottomonium state, provides a particularly clean environment for probing doubly heavy tetraquarks through its weak decays. In Ref.~\cite{Chen:2021erj}, the production of charmed hadronic molecules in $B$-meson decays was analyzed using current algebra and QCD sum rules. Similarly, the $T_{bc\bar{s}}$ states can be accessed via the weak decay of $\Upsilon$, where the $\bar{b}$ quark undergoes a Cabibbo-favored transition $\bar{b} \to \bar{c} + c \bar{s}$, accompanied by the creation of a light quark-antiquark pair from the vacuum. This process yields a final state with quark content $b \bar{c} c \bar{s} q \bar{q}$, where $q$ denotes light quarks. Given the molecular interpretation of $T_{bc\bar{s}}$ arising from $D_s^{(*)+} \bar{B}^{(*)0}$ and $D^{(*)+} \bar{B}_s^{(*)0}$ interaction, the relevant final-state configuration is $[D_s^{(*)+} \bar{B}^{(*)0}/D^{(*)+} \bar{B}_s^{(*)0}] D^{(*)-}$.  As the $T_{bc\bar{s}}^{VV}$ states are dynamically generated hadronic molecules lying just below the corresponding meson-pair thresholds, they are expected to manifest primarily as near-threshold enhancements in the $ \bar{B}^{*0}D_s^{*+}$ or $ \bar{B}_s^{*0}D^{*+}$ invariant mass spectra. However, the $T_{bc\bar{s}}^{VV}$ states can also be observed indirectly in the $\bar{B}^0 D_s^+$ or $\bar B_s^0 D^+$ invariant mass distribution as distinct peak-like structures. This feature allows us to extract the mass and width of the $T_{bc\bar{s}}^{VV}$ states from the $\bar{B}^0 D_s^+$ or $\bar B_s^0 D^+$ invariant mass spectrum. Therefore, in this work we focus on the decay $\Upsilon \to D^{-} \bar{B}^{0} D_s^{+}$. The $T_{bc\bar{s}}$ states with quantum numbers $0^+$ and $2^+$ are expected to appear as enhancements in the $\bar{B}^0 D_s^+$ invariant mass distribution, while the $1^+$ states would not appear in this channel.

A key advantage of this channel is the absence of known resonance structures in the $D^{-} \bar{B}^0$ and $D^{-} D_s^+$ invariant mass spectra. The kinematic limits for this decay are: $7248\,\text{MeV} < m_{\bar{B}^0 D_s^+} < 7591\,\text{MeV}$, $3838\,\text{MeV} < m_{D^{-} D_s^+} < 4181\,\text{MeV}$, and $7149\,\text{MeV} < m_{D^{-} \bar{B}^0} < 7492\,\text{MeV}$. The $D^{-} D_s^+$ system has quark content $\bar{c} d \bar{s} c$, corresponding to a $Z_{cs}$-like state, which must have quantum numbers $J^P = \text{even}^+$ or $\text{odd}^-$. No well-established resonance state satisfying these criteria has been reported in the Review of Particle Physics (RPP) \cite{ParticleDataGroup:2024cfk}. The $D^{-} \bar{B}^0$ system, with quark content $\bar{c} d \bar{d} b$, could in principle support a $B_c$-like hadron. However, no well-established $B_c$-like state satisfying the required criteria has been identified yet. As discussed in Ref.~\cite{Sakai:2017avl}, such systems may form shallow molecular bound states, but no such state has yet been observed experimentally. Therefore, in the present study we focus on the manifestation of the $T_{bc\bar{s}}$ states in the $\bar{B}^0 D_s^+$ invariant mass distribution and set aside the contributions from the potential resonances in the $D^{-} \bar{B}^0$ and $D^{-} D_s^+$ invariant mass spectra, as their influence in the relevant kinematic regions is expected to be limited.


The $\Upsilon$ meson resides below the open-bottom threshold ($M_{\Upsilon} < 2M_B$), which prevents the decay into $B\bar{B}$ pairs. As a $b\bar{b}$ quarkonium state, its dominant decay mechanism proceeds via annihilation into three gluons, yielding hadronic final states through strong interactions. Electromagnetic decays constitute a subdominant channel \cite{CLEO:1998moj}, while weak decays are further suppressed by both the hierarchy of fundamental interactions and competition with strong and electromagnetic processes. A comparable suppression pattern is observed in charmonium: recent BESIII measurements establish only upper limits on weak decay modes of the $J/\psi$ \cite{BESIII:2025rjn}. Consequently, the experimental observation of $\Upsilon$ weak decays remains a challenge with current detector sensitivity and luminosity. Verification of theoretical predictions for such processes will necessitate future high-statistics experiments at next-generation colliders.

In this work we investigate the weak decay process $\Upsilon \to D^- \bar{B}^0 D_s^+$ to probe the $T_{bc\bar{s}}$ tetraquark states. First, we apply current algebra to model the quark-level dynamics of the $\Upsilon$ weak decay. Next, we incorporate the contribution of the $T_{bc\bar{s}}$ states as intermediate resonances dynamically generated through final-state interactions. Finally, under two sets of parameters, we compute the $\bar{B}^0 D_s^+$ invariant mass distribution in this decay.

The paper is organized as follows: In Sec.~\ref{sec:form} we present the theoretical framework, including the quark-level description of the $\Upsilon$ weak decay and the relevant hadron-level Feynman diagrams. In Sec.~\ref{sec:result} we show and discuss the structures in the invariant mass distributions. Finally, a summary is given in Section~\ref{sec:concl}. The mass, width, and coupling strengths of the $T_{bc\bar{s}}$ states to relevant hadronic channels are evaluated and summarized in Appendix~\ref{sec:app-a}.

\section{Formalism}
\label{sec:form}

In this section we present the theoretical formalism employed in the present study. We begin by describing the process $\Upsilon \to D^{-} \bar{B}^{0} D_s^{+}$ using current algebra. This approach has been successfully applied in Ref.~\cite{Chen:2021erj} to systematically investigate the production of hadronic molecular states such as $D^{(*)} \bar{D}^{(*)}$, $D^{(*)} \bar{K}^{(*)}$, and $D^{(*)} D_s^{(*)-}$ in $B$-meson decays.

\begin{figure}[htb]
    \centering
    \subfigure[]{\includegraphics[scale=0.47]{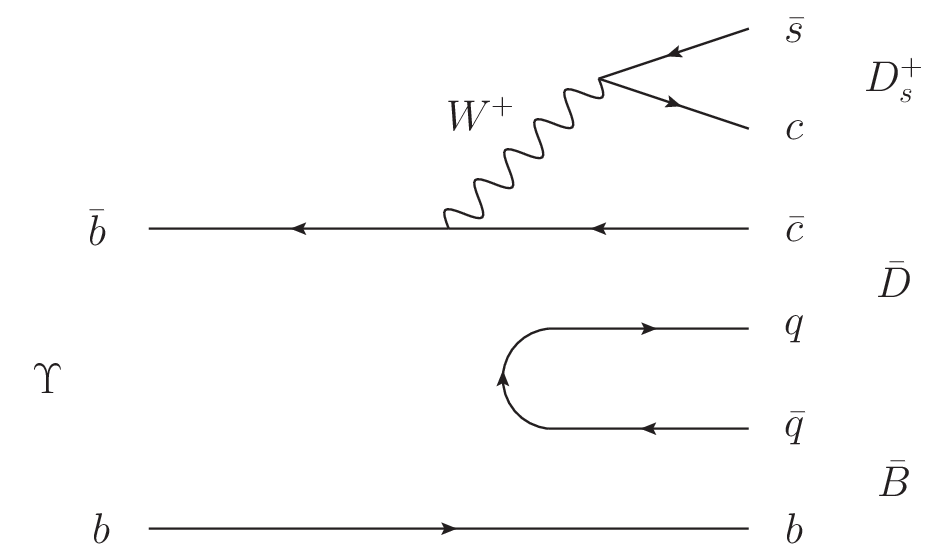}\label{fig:Udecay-quark-01}}
    \quad
    \subfigure[]{\includegraphics[scale=0.47]{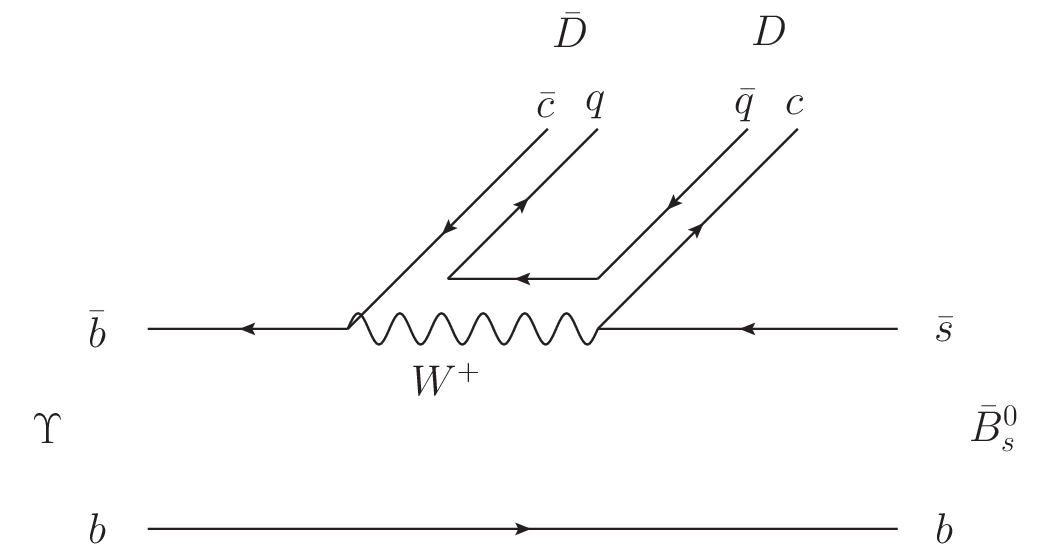}\label{fig:Udecay-quark-02}}
    \caption{Quark-level diagrams for the process $\Upsilon \to b  \bar c + W^+ \to b \bar c  \bar s c  (\bar q q)$, including (a) external emission diagram and (b) internal emission diagram.}
    \label{fig:Udecay}
\end{figure}

As illustrated in Fig.~\ref{fig:Udecay-quark-01}, the initial $\Upsilon$ state has quark content $\bar{b}b$. In its decay, the $\bar{b}$ quark first undergoes a Cabibbo-favored transition $\bar{b} \to \bar{c} + W^+$, with the $W^+$ boson subsequently decaying into a $\bar{s}c$ quark pair. The resulting system $\bar{s}c\bar{c}b$ then hadronizes by incorporating an isoscalar quark-antiquark pair $\bar{u}u + \bar{d}d + \bar{s}s$ from the vacuum. The final-state hadrons are formed accordingly, leading to the following possible configurations:
\begin{eqnarray}
    \Upsilon &=& \bar{b}b \to \bar{s}c\bar{c}b \to \bar{s}c\bar{c}(u\bar u + d\bar d + s\bar s)b \nonumber \\
    &\to& D_s^{(*)+} \bar{D}^{(*)0} B^{(*)-} + D_s^{(*)+} D^{(*)-} \bar{B}^{(*)0} + D_s^{(*)+} D_s^{(*)-} \bar{B}_s^{(*)0} + \cdots.
\end{eqnarray}

Thus, the tetraquark molecular states $T_{bc\bar{s}}$ can naturally manifest in the final states $D_s^{(*)+} \bar{D}^{(*)0} B^{(*)-}$ or $D_s^{(*)+} D^{(*)-} \bar{B}^{(*)0}$.  In this study we focus on the $T_{bc\bar{s}}$ states with quark content $b c \bar{s} \bar{d}$, which corresponds specifically to the final state $[D_s^{(*)+} \bar{B}^{(*)0}] D^{(*)-}$. To maximize the phase space and thereby obtain as many events experimentally as possible, we choose the decay final state to be $[D_s^+ \bar B^0] D^-$. We also consider the rescattering contribution via the intermediate process $\Upsilon \to D^{-} \bar{B}^{*0} D_s^{*+} \to D^{-} \bar{B}^{0} D_s^{+}$. This is motivated by the fact that the $T_{bc\bar{s}}^{VV}$ states, dynamically generated from vector-vector interactions, strongly couple to the $\bar{B}^* D_s^*$ and $\bar{B}_s^* D^*$ channels. Including the $\Upsilon \to D^{-} \bar{B}^{*0} D_s^{*+}$ transition is therefore essential for properly accounting for the influence of these intermediate states on the $\bar{B}^0 D_s^+$ invariant mass distribution.

The internal emission mechanism depicted in Fig.~\ref{fig:Udecay-quark-02} can also produce the $[D^{(*)+} \bar{B}_s^{(*)0}] D^{(*)-}$ final state, where $D^{(*)+} \bar{B}_s^{(*)0}$ forms an important component of the $T_{bc\bar{s}}$ molecular states with the same quark content $b c \bar{s} \bar{d}$ \cite{Liu:2023hrz}. This production mechanism involved internal emission leads to a color suppression factor of $1/3$. Therefore, for the sake of simplicity, we consider only the contribution from Fig.~\ref{fig:Udecay-quark-01} in this work.

We shall develop the Fierz rearrangement in Sec.~\ref{sec:fierz} to describe the production mechanisms depicted in Fig.~\ref{fig:Udecay-quark-01}, and further consider the production of the $T_{bc \bar s}$ states through final-state interactions, ultimately presenting the dominant Feynman diagrams for the process $\Upsilon \to D^{-} \bar{B}^{0} D_s^{+}$ in Sec.~\ref{sec:amp}.

\subsection{Fierz Rearrangement}
\label{sec:fierz}

To describe the production mechanism depicted in Fig.~\ref{fig:Udecay-quark-01}, we employ the color rearrangement identity $\delta_{ab} \delta_{ef} = \frac{1}{3} \delta_{af} \delta_{eb} + \frac{1}{2} \lambda_{af}^n \lambda_{eb}^n$, and apply the Fierz transformation to exchange the quarks $q_f$ and $b_b$ in the Dirac spinor space:
\begin{align}
   \Upsilon \longrightarrow  & \quad J_{\Upsilon} =  [\delta^{ab} \bar b_a \gamma_\mu b_b] \\
    \stackrel{weak}{\longrightarrow}& \quad [\delta^{ab} \bar c_a  \gamma_\rho  (1-\gamma_5) \gamma_\mu  b_b] \times [\delta^{cd} \bar s_c \gamma^\rho (1-\gamma_5) c_d] 
    \label{eq:fierz-weak} \\
    \stackrel{QPC}{\longrightarrow}& \quad [\delta^{ab} \bar c_a   \gamma_\rho (1-\gamma_5) \gamma_\mu  b_b] \times [\delta^{cd} \bar s_c \gamma^\rho (1-\gamma_5) c_d] \times [\delta^{ef} \bar q_e q_f] 
    \label{eq:fierz-qpc} \\
    \stackrel{color}{\longrightarrow}& \quad {{ \delta^{af} \delta^{eb} \delta^{cd} }\over{ 3 }} \times [ \bar c_a  \gamma_\rho  (1-\gamma_5) \gamma_\mu  b_b] \times [ \bar s_c \gamma^\rho (1-\gamma_5) c_d] \times [ \bar q_e q_f] + \cdots 
    \label{eq:fierz-color} \\ \nonumber
    \stackrel{Fierz:b_b \leftrightarrow q_f}{\longrightarrow}
      & \quad -\frac{1}{12} \times [ \delta^{af} \bar c_a \gamma_\rho (1-\gamma_5) q_f]  \times [\delta^{eb} \bar q_e \gamma_\mu (1+\gamma_5) b_b] \times  [\delta^{cd} \bar s_c \gamma_\rho (1-\gamma_5) c_d] \\ \nonumber
      &\quad -\frac{1}{12} \times [ \delta^{af} \bar c_a   (1+\gamma_5) q_f]  \times [\delta^{eb} \bar q_e (1+\gamma_5)  b_b] \times  [\delta^{cd} \bar s_c \gamma_\mu (1-\gamma_5) c_d] \\ \nonumber
      &\quad -\frac{i}{12} \times [ \delta^{af} \bar c_a  \sigma_{\mu\rho} (1+\gamma_5) q_f]  \times [\delta^{eb} \bar q_e (1+\gamma_5)  b_b] \times  [\delta^{cd} \bar s_c \gamma_\rho (1-\gamma_5) c_d] \\ \nonumber
      &\quad -\frac{i}{12} \times [ \delta^{af} \bar c_a  (1+\gamma_5) q_f]  \times [\delta^{eb} \bar q_e \sigma_{\mu\rho} (1+\gamma_5)  b_b] \times  [\delta^{cd} \bar s_c \gamma_\rho (1-\gamma_5) c_d] \\ \nonumber
      &\quad +\frac{1}{12} \times [ \delta^{af} \bar c_a \sigma_{\nu\rho} (1+\gamma_5) q_f]  \times [\delta^{eb} \bar q_e \sigma_{\mu\nu} (1+\gamma_5)  b_b] \times  [\delta^{cd} \bar s_c \gamma_\rho (1-\gamma_5) c_d] \\ \nonumber
      &\quad +\frac{i}{12} \times [ \delta^{af} \bar c_a \gamma_l  (1-\gamma_5) q_f]  \times [\delta^{eb} \bar q_e \gamma_\nu(1+\gamma_5)  b_b] \times  [\delta^{cd} \bar s_c \gamma_\rho (1-\gamma_5) c_d] \, \epsilon_{\mu \nu \rho l} \\ \nonumber
      &\quad -\frac{1}{12} \times [ \delta^{af} \bar c_a  \gamma_\nu (1-\gamma_5) q_f]  \times [\delta^{eb} \bar q_e \gamma_\nu (1+\gamma_5)  b_b] \times  [\delta^{cd} \bar s_c \gamma_\mu (1-\gamma_5) c_d]\\ \nonumber
      &\quad +\frac{1}{12} \times [ \delta^{af} \bar c_a  \gamma_\mu (1-\gamma_5) q_f]  \times [\delta^{eb} \bar q_e \gamma_\rho (1+\gamma_5)  b_b] \times  [\delta^{cd} \bar s_c \gamma_\rho (1-\gamma_5) c_d] \nonumber \\ 
      &\quad    +  \cdots . 
      \label{eq:fierz}
\end{align}

A brief explanation of the Fierz rearrangement in Eq.~(\ref{eq:fierz}) is as follows. Eq.~(\ref{eq:fierz-weak}) describes the Cabibbo-favored weak decay process $\bar{b} \to \bar{c} + \bar{s}c$ through the $V-A$ current. Eq.~(\ref{eq:fierz-qpc}) accounts for the creation of a quark-antiquark pair $\bar{q}q$ from the vacuum via the $^3P_0$ quark pair creation mechanism. Eq.~(\ref{eq:fierz-color}) applies the color rearrangement identity $\delta_{ab}\delta_{ef} = \frac{1}{3}\delta_{af}\delta_{eb} + \frac{1}{2}\lambda_{af}^n \lambda_{eb}^n$ to project the system into the appropriate color-singlet configurations. The $\cdots$ represents contributions from the color-octet configurations, which are neglected in the present work. Finally, Eq.~(\ref{eq:fierz}) incorporates the Fierz transformation to interchange the $q_f$ and $b_b$ quarks in Dirac spinor space, thereby placing the quarks into the current operators corresponding to the appropriate hadronic final states.

\subsection{Decay Amplitude}
\label{sec:amp}

In the previous subsection we applied Fierz rearrangement to describe the quark-level mechanism corresponding to Fig.~\ref{fig:Udecay-quark-01}, which allows the resulting quark-level configuration from the weak transition to be expressed in terms of a set of local quark current operators listed in Eq.~(\ref{eq:fierz}). By matching the quark content and quantum numbers of the current operators to the desired final-state hadrons, $D^{-} \bar{B}^{0} D_s^{+}$ and $D^{-} \bar{B}^{*0} D_s^{*+}$, we can establish a direct connection between the quark-level dynamics and the physical hadronic decay process.

Pseudoscalar mesons $D$, $D_s$, and $B$ couple to both axial-vector and pseudoscalar currents, as described by the matrix elements:
\begin{align}
    \langle 0 | \bar{q} \gamma_\mu \gamma_5 q^\prime | P(q) \rangle &= i q_\mu f_P \,,
    \label{eq:coup-A} \\
    \langle 0 | \bar{q} i \gamma_5 q^\prime | P(q) \rangle &= \lambda_P \,,
    \label{eq:coup-P}
\end{align}
where $f_D = 212\,\text{MeV}$ \cite{ParticleDataGroup:2024cfk}, $f_{D_s} = 250\,\text{MeV}$ \cite{ParticleDataGroup:2024cfk}, and $f_B = 190\,\text{MeV}$ \cite{ParticleDataGroup:2024cfk} are the meson decay constants. The parameter $\lambda_P$ is related to the decay constant and meson mass via $\lambda_D = f_D m_D^2 / m_c$ and $\lambda_B = f_B m_B^2 / m_b$, with the current quark masses taken as $m_c = 1270\,\text{MeV}$ and $m_b = 4180\,\text{MeV}$ \cite{ParticleDataGroup:2024cfk}.

Vector mesons $D_s^*$ and $B^*$ couple to vector and tensor currents through the following matrix elements:
\begin{align}
    \langle 0 | \bar{q} \gamma_\mu q^\prime | V(\epsilon, q) \rangle &= m_V f_V \epsilon_\mu \,, 
    \label{eq:coup-V} \\
    \langle 0 | \bar{q} \sigma_{\mu\nu} q^\prime | V(\epsilon, q) \rangle &= i f_V^T (q_\mu \epsilon_\nu - q_\nu \epsilon_\mu) \,,
    \label{eq:coup-T1} \\
    \langle 0 | \bar{q} \sigma_{\mu\nu} \gamma_5 q^\prime | V(\epsilon, q) \rangle &= -\frac{1}{2} \epsilon_{\mu\nu\alpha\beta} f_V^T (q_\alpha \epsilon_\beta - q_\beta \epsilon_\alpha) \,,
    \label{eq:coup-T2}
\end{align}
where $f_{D_s^*} = 301\,\text{MeV}$ \cite{ParticleDataGroup:2020ssz} and $f_{B^*} = 213\,\text{MeV}$ \cite{Wang:2015mxa}. For the tensor decay constant $f_{B^*}^T$, we use the approximation $f_{B^*}^T \simeq f_{B^*} \cdot (f_{J/\psi}^T / f_{J/\psi})$, with $f_{J/\psi}^T = 410\,\text{MeV}$ and $f_{J/\psi} = 418\,\text{MeV}$ \cite{Becirevic:2013bsa}. These couplings allow the quark-level currents to be matched onto physical hadronic states, enabling the calculation of decay amplitudes in terms of measurable hadronic parameters.

By combining the current operators with the hadronic coupling formulas given in Eqs.~(\ref{eq:coup-A})-(\ref{eq:coup-T2}), we extract from Eq.~(\ref{eq:fierz}) the direct decay amplitude for the process $\Upsilon \to D^{-} \bar{B}^{0} D_s^{+}$:
\begin{eqnarray}
&& \langle \Upsilon(\epsilon_0,p) | \bar{D}(p_3)~\bar{B}(p_2)~D_s^{+}(p_1) \rangle
\nonumber \\
&=&     {i\, a_1 \over 12} \, f_D f_{B} f_{D_s} \mathcal{A}_1
 -    {i\, a_1 \over 12} \, \lambda_D \lambda_B f_{D_s} \, \epsilon_0 \cdot p_1 \,, 
 \label{eq:me-pp} \\
\mathcal{A}_1 &=& p_1^\rho p_2^\nu (\epsilon_{0\rho}p_{3\nu} + \epsilon_{0\nu}p_{3\rho} - g_{\rho\nu} \epsilon_0 \cdot p_3 - i \epsilon_{\mu\nu\rho l} \, \epsilon_0^\mu  p_3^l ) \, ,
\label{eq:amp-A1}
\end{eqnarray}
as well as the amplitude for the vector-vector final state $\Upsilon \to D^{-} \bar{B}^{*0} D_s^{*+}$:
\begin{eqnarray}
&& \langle \Upsilon(\epsilon_0,p) | \bar{D}(p_3)~\bar{B}^*(\epsilon_2,p_2)~D_s^{*+}(\epsilon_1,p_1) \rangle
\nonumber \\
&=&    {i\, a_1 \over 12} \, f_D f_{B^*} f_{D_s^*} m_{B^*} m_{D_s^*} \mathcal{A}_2  -    {i \, a_1 \over 12} \, \lambda_D f_{B^*}^T f_{D_s^*} m_{D_s^*} \mathcal{A}_3  \,,
\label{eq:me-vv} \\
\mathcal{A}_2 &=& \left( \epsilon_1 \cdot p_3 \, \epsilon_2 \cdot \epsilon_0 + \epsilon_2 \cdot p_3 \, \epsilon_0 \cdot \epsilon_1 - \epsilon_0 \cdot p_3 \, \epsilon_1 \cdot \epsilon_2 - i \epsilon_{\mu\nu\rho\sigma} \, \epsilon_0^\mu \epsilon_2^\nu \epsilon_1^\rho p_3^\sigma \right) \, ,
\label{eq:amp-A2} \\
\mathcal{A}_3 &=& \left( q_2^\mu \epsilon_2^\rho - q_2^\rho \epsilon_2^\mu + {i\over2} \epsilon_{\mu\rho\alpha\beta} (q_2^\alpha \epsilon_2^\beta - q_2^\beta \epsilon_2^\alpha) \right) \epsilon_0^\mu \epsilon_1^\rho \, .
\label{eq:amp-A3}
\end{eqnarray}
In the equation, we introduce an overall factor $a_1$, which incorporates contributions from three distinct sources: (a) the coupling between $\Upsilon$ and the current $J_\Upsilon$, (b) the coupling strength of the weak interaction process, and (c) the probability of quark-antiquark pair creation from the vacuum via the $^3P_0$ mechanism.

\begin{figure}[htb]
    \centering
    \subfigure[]{\includegraphics[scale=0.5]{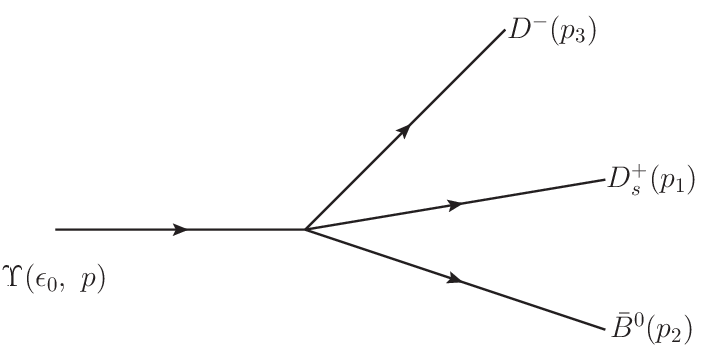}\label{fig:Udecay-a}} \quad\quad
    \subfigure[]{\includegraphics[scale=0.5]{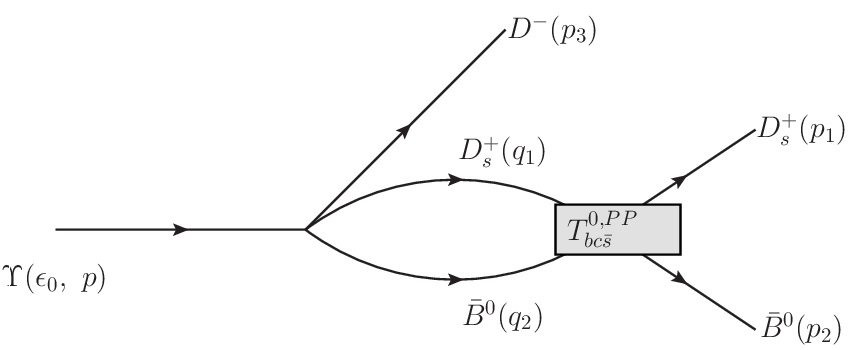}\label{fig:Udecay-b}}\\
    \subfigure[]{\includegraphics[scale=0.5]{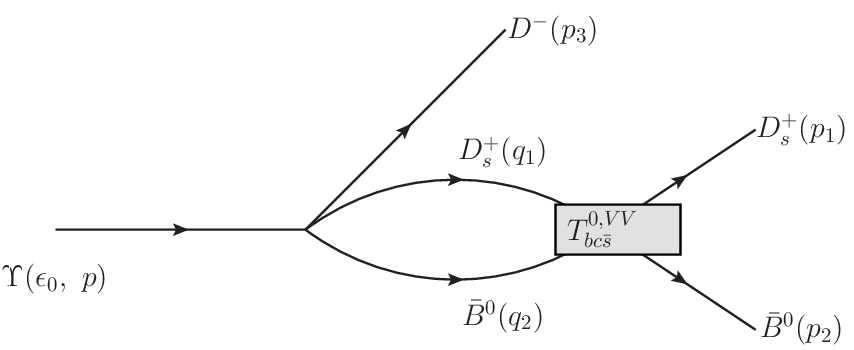}\label{fig:Udecay-c}} \quad\quad
    \subfigure[]{\includegraphics[scale=0.5]{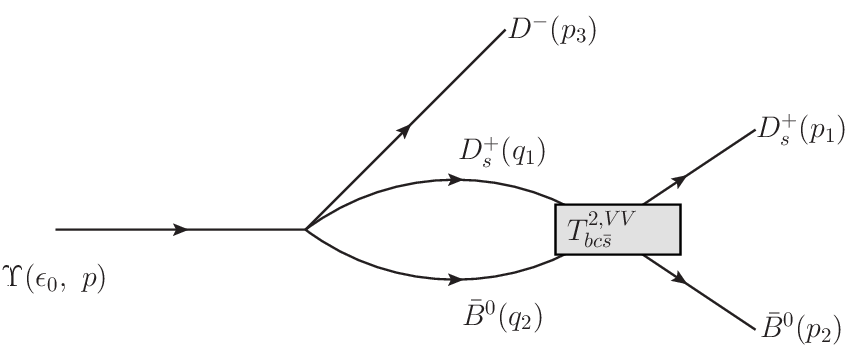}\label{fig:Udecay-d}}\\
    \subfigure[]{\includegraphics[scale=0.5]{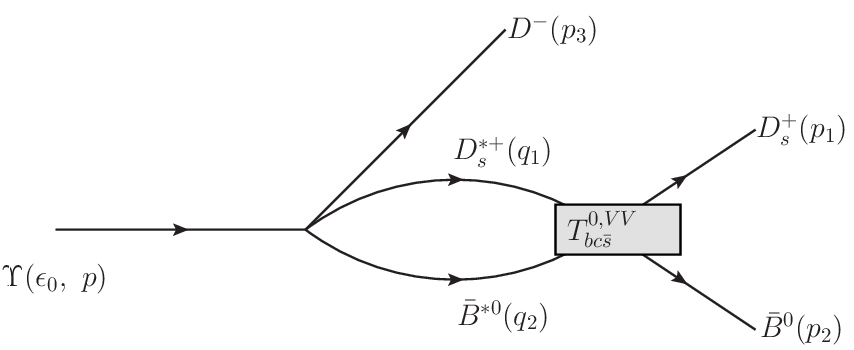}\label{fig:Udecay-e}} \quad\quad
    \subfigure[]{\includegraphics[scale=0.5]{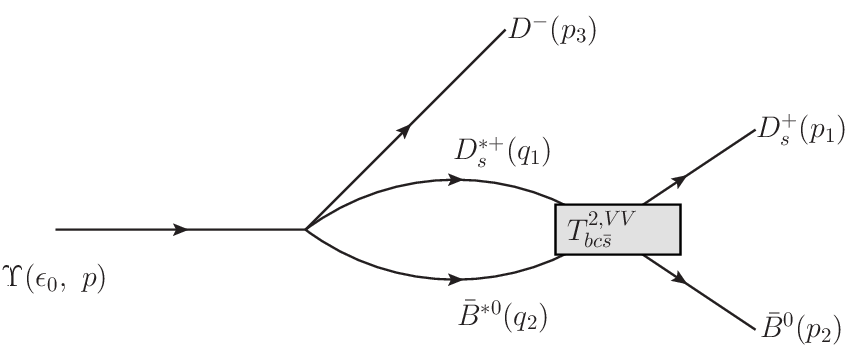}\label{fig:Udecay-f}}
    \caption{Feynman diagrams for the process $\Upsilon \to D^{-} \bar{B}^{0} D_s^{+}$. (a) Direct production via the tree-level mechanism. (b), (c), and (d): Rescattering contributions mediated by the intermediate states $T_{bc \bar s}^{0, PP}$, $T_{bc \bar s}^{0, VV}$, and $T_{bc \bar s}^{2, VV}$ in the $\bar{B}^0 D_s^+$ final state, respectively. (e) and (f): Contributions from $T_{bc \bar s}^{0, VV}$ and $T_{bc \bar s}^{2, VV}$ via the $\bar{B}^{*0} D_s^{*+}$ intermediate channel, which subsequently undergoes hadronic transitions to the final $\bar{B}^0 D_s^+$ state.}
    \label{fig:Udecay-amp}
\end{figure}

We can now construct the Feynman diagrams representing the dominant contributions to the process $\Upsilon \to D^{-} \bar{B}^{0} D_s^{+}$, as illustrated in Fig.~\ref{fig:Udecay-amp}. In this framework we include the states $T_{bc \bar{s}}^{0, PP}$, $T_{bc \bar{s}}^{0, VV}$, and $T_{bc \bar{s}}^{2, VV}$ as intermediate resonances. Specifically, $T_{bc \bar{s}}^{0, PP}$ denotes the scalar resonance dynamically generated from the S-wave interaction of $\bar{B}_s^0 D^+ \,\text{--}\, \bar{B}^0 D_s^+$ system, while $T_{bc \bar{s}}^{0, VV}$ and $T_{bc \bar{s}}^{2, VV}$ correspond to the $J^P = 0^+$ and $2^+$ resonances arising from the S-wave interaction of $\bar{B}_s^{*0} D^{*+} \,\text{--}\, \bar{B}^{*0} D_s^{*+}$ system, respectively. The properties of these states have been investigated in our previous work \cite{Liu:2023hrz} and are further detailed in Appendix~\ref{sec:app-a}.

The tree-level decay amplitude corresponding to Fig.~\ref{fig:Udecay-a} and derived from Eq.~\eqref{eq:me-pp} is given by:
\begin{align}
    i\mathcal{M}^{a}_{tree} =& C_1 ~ \mathcal{A}_1 -  C_2 ~ \epsilon_0 \cdot p_1 \, , 
    \label{eq:amp-a} 
\end{align}
where $ C_1 = { a_1 \over 12} ~ f_D f_{B} f_{D_s}$ and $ C_2 = { a_1 \over 12}~ \lambda_D \lambda_B f_{D_s}$.

The rescattering contributions via the intermediate $T_{bc\bar{s}}$ states in the $\bar{B}^0 D_s^+$ channel are described by the loop amplitudes:
\begin{align}
    i\mathcal{M}^{b}_{loop} =& i \int {d^4 q_1\over{(2\pi)^4}}{{ C_1 ~ \mathcal{A}_4 -  C_2 ~ \epsilon_0 \cdot q_1 }\over (q_1^2 - m_1^2 + i \epsilon)((P_{12} - q_1)^2 - m_2^2 + i \epsilon)} ~ {{g^2_{X_1}}\over{m_{12}^2-w_{X_1}^2}} 
    \label{eq:amp-b} \, , \\
    i\mathcal{M}^{c}_{loop} =& i \int {d^4 q_1\over{(2\pi)^4}}{{ C_1 ~ \mathcal{A}_4 -  C_2 ~ \epsilon_0 \cdot q_1 }\over (q_1^2 - m_1^2 + i \epsilon)((P_{12} - q_1)^2 - m_2^2 + i \epsilon)} ~ {{g^2_{X_2}}\over{m_{12}^2-w_{X_2}^2}} 
    \label{eq:amp-c} \, , \\
    i\mathcal{M}^{d}_{loop} =& i  \int {d^4 q_1\over{(2\pi)^4}}{{ ( C_1 ~ \mathcal{A}_4 -  C_2 ~ \epsilon_0 \cdot q_1) q_1^\mu q_2^\nu }\over (q_1^2 - m_1^2 + i \epsilon)((P_{12} - q_1)^2 - m_2^2 + i \epsilon)}    ~ {{g^2_{X_3}}\over{m_{12}^2-w_{X_3}^2}} \mathcal{P}^2_{\mu\nu\rho\sigma}  p_1^\rho p_2^\sigma \, , 
    \label{eq:amp-d} \\
    \mathcal{A}_4 =& p_1^\rho p_2^\nu (\epsilon_{0\rho}p_{3\nu} + \epsilon_{0\nu}p_{3\rho} - g_{\rho\nu} \epsilon_0 \cdot p_3 - i \epsilon_{\mu\nu\rho l} \, \epsilon_0^\mu  p_3^l ) \, ,
    \label{eq:amp-PA2}
\end{align}
where $\mathcal{P}^2_{\mu\nu\rho\sigma} = \frac{1}{2} (\mathcal{P}^1_{\mu\rho} \mathcal{P}^1_{\nu\sigma} + \mathcal{P}^1_{\mu\sigma} \mathcal{P}^1_{\nu\rho}) - \frac{1}{3} \mathcal{P}^1_{\mu\nu} \mathcal{P}^1_{\rho\sigma}$ is the spin-2 projection operator, with $\mathcal{P}^1_{\mu\nu} = -g_{\mu\nu} + q_\mu q_\nu / q^2$. For these loop amplitudes, the contributions from the $T_{bc\bar{s}}$ intermediate states are introduced through the effective Lagrangians in Eqs.~(\ref{eq:L-coup-PPX0}) and (\ref{eq:L-coup-PPX2}).

As previously discussed, the final state $D^{-} \bar{B}^{0} D_s^{+}$ can also be reached through rescattering via the vector intermediate state $\bar{B}^{*0} D_s^{*+}$, particularly when the amplitude for $\Upsilon \to D^{-} \bar{B}^{*0} D_s^{*+}$ is comparable to the direct tree-level process for $\Upsilon \to D^{-} \bar B^{0} D_s^{+}$. This mechanism enhances the sensitivity to the $T_{bc\bar{s}}^{0,VV}$ and $T_{bc\bar{s}}^{2,VV}$ resonances, which couple strongly to the vector-vector channel $\bar{B}^{*0} D_s^{*+}$. We therefore include the loop amplitudes for the diagrams in Figs.~\ref{fig:Udecay-e} and \ref{fig:Udecay-f}:
\begin{align}
    i\mathcal{M}^{e}_{loop} =&  i \int {d^4 q_1\over{(2\pi)^4}}   { \mathcal{A}_5 (-g_{\alpha \gamma} + {q_{1 \alpha} q_{1 \gamma} \over m_1^2 })  (-g_{\beta \gamma} + {q_{2 \beta} q_{2 \gamma} \over m_2^2 }) \over (q_1^2 - m_1^2 + i \epsilon)((P_{12} - q_1)^2 - m_2^2 + i \epsilon)} ~ {{{1 \over \sqrt{3}} g^\prime_{X_2} g_{X_2}}\over{m_{12}^2-w_{X_2}^2}} \, , 
    \label{eq:amp-e} \\
    i\mathcal{M}^{f}_{loop} =& i \int {d^4 q_1\over{(2\pi)^4}}   { \mathcal{A}_5 (-g_{\alpha \mu} + {q_{1 \alpha} q_{1 \mu} \over m_1^2 })  (-g_{\beta \nu} + {q_{2 \beta} q_{2 \nu} \over m_2^2 }) \over (q_1^2 - m_1^2 + i \epsilon)((P_{12} - q_1)^2 - m_2^2 + i \epsilon)} ~ {{ g^\prime_{X_3} g_{X_3}}\over{m_{12}^2-w_{X_3}^2}} \mathcal{P}^2_{\mu\nu\rho\sigma} p_1^\rho p_2^\sigma \, ,
    \label{eq:amp-f} \\
    \mathcal{A}_5 =& C_3 (p_3^\alpha \epsilon_0^\beta + p_3^\beta \epsilon_0^\alpha - \epsilon_0 \cdot p_3 g^{\alpha \beta} )  - C_4 ( \epsilon_0 \cdot q_2 g^{\alpha\beta} -q_2^\alpha \epsilon_0^\beta ) \, ,
    \label{eq:amp-PA3}
\end{align}
where $ C_3 = { a_1 \over 12}~ f_D f_{B^*} f_{D_s^*} m_{B^*} m_{D_s^*}$ and $ C_4 = { a_1 \over 12}~  \lambda_D f_{B^*}^T f_{D_s^*} m_{D_s^*} $. It is worth noting that a special factor of $\frac{1}{\sqrt{3}}$ appears in Eq.~(\ref{eq:amp-e}), which originates from the normalization choice made in our calculation of the coupling constants in Appendix~\ref{sec:app-a}. For the coupling of $T_{bc\bar{s}}^{0,VV}$ and $T_{bc\bar{s}}^{2,VV}$ states to the vector-vector states ($VV$), we adopt the following form:  
\begin{align}
    \mathcal{L}_{V V T_{bc\bar{s}}^{0,VV}} &= \frac{1}{\sqrt{3}} g \, V^\mu V_\mu T_{bc\bar{s}}^{0} \, , \\
    \mathcal{L}_{V V T_{bc\bar{s}}^{2,VV}} &=  g \, V_\mu V_\nu T_{bc\bar{s}}^{2,\mu\nu} \, .
\end{align}

In Eqs.~\eqref{eq:amp-a}–\eqref{eq:amp-PA3}, the $w_{X_1}$, $w_{X_2}$, and $w_{X_3}$ appearing in the denominator of the Breit-Wigner form are the pole positions of the $T_{bc\bar{s}}^{0,PP}$, $T_{bc\bar{s}}^{0,VV}$, and $T_{bc\bar{s}}^{2,VV}$ states, respectively. The coupling constants $g_{X_1}$, $g_{X_2}$, and $g_{X_3}$ represent the couplings of $T_{bc\bar{s}}^{0,PP}$, $T_{bc\bar{s}}^{0,VV}$, and $T_{bc\bar{s}}^{2,VV}$ to the $\bar{B}^0 D_s^+$ channel, respectively. The couplings $g'_{X_2}$ and $g'_{X_3}$ denote the corresponding couplings to the $\bar{B}^{*0} D_s^{*+}$ channel. All these pole positions and couplings are determined in Appendix~\ref{sec:app-a} based on the molecular picture.

Another crucial aspect of our analysis is the regularization of the loop diagrams. The amplitudes $\mathcal{M}^{b}_{\text{loop}}$, $\mathcal{M}^{c}_{\text{loop}}$, $\mathcal{M}^{d}_{\text{loop}}$, $\mathcal{M}^{e}_{\text{loop}}$, and $\mathcal{M}^{f}_{\text{loop}}$ involve loop integrals that are ultraviolet divergent and must be regularized to yield physically meaningful results.

A commonly used approach is the three-momentum cutoff scheme (CO), which was employed in our previous study \cite{Liu:2023hrz} to determine the pole positions and coupling strengths of the $T_{bc\bar{s}}$ states. In this scheme, the loop integral is evaluated by first integrating over the energy component $q_0$, followed by imposing a cutoff $q_{\text{max}}$ on the three-momentum magnitude: $|\vec{q}| < q_{\text{max}}$.

However, to better preserve the underlying field-theoretic structure, we adopt dimensional regularization (DR) in this work. In order to maintain consistency with the results of Ref.~\cite{Liu:2023hrz}, we determine the subtraction constants such that the finite part of the loop amplitude at threshold matches that obtained with the three-momentum cutoff scheme. Specifically, we impose the condition
\begin{align}
    \mathcal{M}^{\text{finite}}_{\text{DR}}(s) \big|_{s = (m_1 + m_2)^2} = \mathcal{M}^{\text{finite}}_{\text{CO}} \big|_{s = (m_1 + m_2)^2} \,,
    \label{eq:matchDR}
\end{align}
where $\mathcal{M}^{\text{finite}}_{\text{CO}} \big|_{s = (m_1 + m_2)^2}$ denotes the finite amplitude in the CO scheme evaluated at threshold. This matching procedure ensures consistency with our earlier determination of the properties of the $T_{bc\bar{s}}$ states, while extending the validity of the amplitude to a broader energy range.





\subsection{Invariant Mass Distribution}

The total amplitude for the process $\Upsilon \to D^{-} \bar{B}^{0} D_s^{+}$ is constructed by summing the contributions from both the direct tree-level decay and the rescattering mechanisms mediated by intermediate resonance states. It is given by:
\begin{align}
    \mathcal{M}_{\text{total}} = \mathcal{M}^{a}_{\text{tree}} + \mathcal{M}^{b}_{\text{loop}} + \mathcal{M}^{c}_{\text{loop}} + \mathcal{M}^{d}_{\text{loop}} + \mathcal{M}^{e}_{\text{loop}} + \mathcal{M}^{f}_{\text{loop}} \,,
    \label{eq:total_amp}
\end{align}
where $\mathcal{M}^{a}_{\text{tree}}$ represents the direct production amplitude, while the loop amplitudes $\mathcal{M}^{b}_{\text{loop}}$–$\mathcal{M}^{f}_{\text{loop}}$ account for the rescattering effects via intermediate $T_{bc\bar{s}}$ states in both the $\bar{B}^0 D_s^+$ and $\bar{B}^{*0} D_s^{*+}$ channels, as derived in the previous section. These contributions collectively encode the dynamical effects of the $T_{bc\bar{s}}^{0,PP}$, $T_{bc\bar{s}}^{0,VV}$, and $T_{bc\bar{s}}^{2,VV}$ resonances, allowing us to probe their imprints in the invariant mass spectrum.

The differential decay width for a $1 \to 3$ body process is given by the standard phase-space formula:
\begin{align}
    \frac{d\Gamma}{dm_{12}^2\, dm_{23}^2} = \frac{1}{(2\pi)^3} \frac{1}{32 M_{\Upsilon}^3} \left| \bar{\mathcal{M}}_{\text{total}} \right|^2 \,.
    \label{eq:diff_width}
\end{align}

\section{Results and discussions}
\label{sec:result}


In this section we perform numerical calculations using two distinct schemes, corresponding to the two sets of parameters determined in Appendices~\ref{sec:app-a-1} and \ref{sec:app-a-2}, respectively. These parameters include the pole positions of the potential $T_{bc\bar{s}}$ molecular states and their coupling strengths to the relevant hadronic channels.

In Scheme I, the finite part of the loop amplitude is obtained by matching the dimensional regularization results to the three-momentum cutoff scheme at threshold, as specified by Eq.~\eqref{eq:matchDR}. To maintain consistency with the analysis in Appendix~\ref{sec:app-a-1}, the three-momentum cutoff is set to $q_{\text{max}} = 450\,\text{MeV}$ in that scheme. The pole positions and coupling constants of the $T_{bc\bar{s}}$ states used in this scheme are taken from Eq.~\eqref{eq:coupling-I} and Table~\ref{tab:results1} in Appendix~\ref{sec:app-a-1}.

\begin{figure}[htb]
    \centering
    \subfigure[]{\includegraphics[scale=0.8]{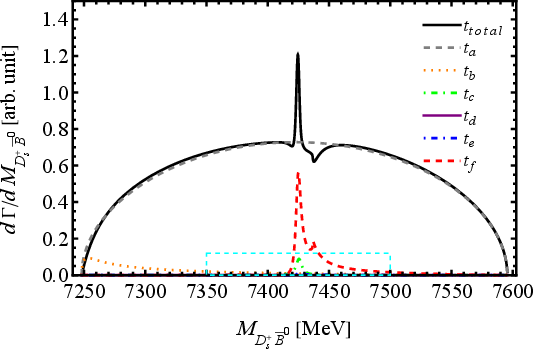}\label{fig:ls-450-a}}
    \quad\quad
    \subfigure[]{\includegraphics[scale=0.8]{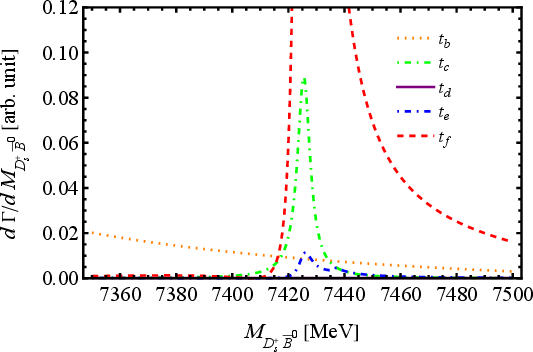}\label{fig:ls-450-b}}
    \caption{(a) The $\bar{B}^0 D_s^+$ invariant mass distribution for the process $\Upsilon \to D^{-} \bar{B}^{0} D_s^{+}$, calculated using the parameters from Scheme I. A clear peak structure is observed near $7425\,\text{MeV}$, which corresponds to the contribution of the $T_{bc\bar{s}}^{2,VV}$ state. (b) The region of (a) enclosed by the cyan dashed line, highlighting the relative magnitudes of the individual amplitudes $t_b, t_c, \dots, t_f$. The legend entries $t_a, t_b, \dots, t_f$ represent the partial contributions from the amplitudes $\mathcal{M}^{a}_{\text{tree}}$, $\mathcal{M}^{b}_{\text{loop}}$, \dots, $\mathcal{M}^{f}_{\text{loop}}$, respectively.}
    \label{fig:ls-450}
\end{figure}

In Fig.~\ref{fig:ls-450}, we show the $\bar{B}^0 D_s^+$ invariant mass distribution. As seen in Fig.~\ref{fig:ls-450-a}, the overall shape of the distribution is smooth, with two notable structures: a sharp peak near $7425\,\text{MeV}$ and a dip located at the $\bar{B}^{*0} D_s^{*+}$ threshold.

The dominant contribution comes from the tree-level diagram, denoted as $t_a$. This explains the generally smooth behavior of the spectrum. The peak at $7425\,\text{MeV}$ is primarily due to the $T_{bc\bar{s}}^{2,VV}$ state. Specifically, the rescattering process $\Upsilon \to D^{-} \bar{B}^{*0} D_s^{*+} \to D^{-} T_{bc\bar{s}}^{2,VV} \to D^{-} \bar{B}^{0} D_s^{+}$, represented by the red dashed line $t_f$, gives a significantly larger contribution. The dip at the $\bar{B}^{*0} D_s^{*+}$ threshold arises from the interplay between the enhanced rescattering amplitude $t_f$ and the tree-level amplitude $t_a$.

In comparison, the amplitude $t_c$, corresponding to $\Upsilon \to D^{-} \bar{B}^{0} D_s^{+} \to D^{-} T_{bc\bar{s}}^{0,VV} \to D^{-} \bar{B}^{0} D_s^{+}$, is about five times smaller than $t_f$. Since $T_{bc\bar{s}}^{0,VV}$ has a mass and width very close to those of $T_{bc\bar{s}}^{2,VV}$ (see Table~\ref{tab:results1}), its individual contribution is difficult to resolve in the spectrum. We also examine the contribution from $t_b$, which corresponds to the process $\Upsilon \to D^{-} \bar{B}^{0} D_s^{+} \to D^{-} T_{bc\bar{s}}^{0,PP} \to D^{-} \bar{B}^{0} D_s^{+}$. While $t_b$ is sizable near threshold and comparable in magnitude to $t_a$ in that region, no significant threshold enhancement emerges after interference. This suggests that, for the current parameter choice with $q_{\text{max}} = 450\,\text{MeV}$, the $\bar{B}^0 D_s^+$ component in the $T_{bc\bar{s}}^{0,PP}$ molecular state is relatively weak.

A partial magnification of the spectrum is provided in Fig.~\ref{fig:ls-450-b}, illustrating the relative contributions of different amplitudes. It is evident that contributions from $t_b$, $t_c$, and $t_f$ are the most relevant beyond the tree term, while all other amplitudes are negligible. Notably, the amplitude $t_e$ is smaller than $t_c$, indicating that in the process $\Upsilon \to D^{-} \bar{B}^{*0} D_s^{*+}$, the final state couples more weakly to the spin-0 component of the $\bar{B}^{*0} D_s^{*+}$ system compared to the spin-2 component. This reflects a suppression of the $J=0$ configuration in the $\bar{B}^{*0} D_s^{*+}$ production at $\Upsilon$ weak decay under the present framework.

\begin{figure}[htb]
    \centering
    \subfigure[]{\includegraphics[scale=0.8]{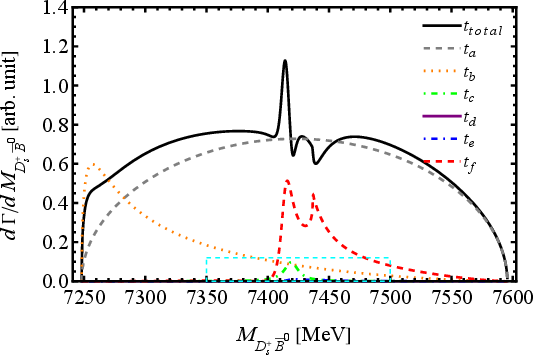}\label{fig:ls-550-a}}
    \quad \quad
    \subfigure[]{\includegraphics[scale=0.8]{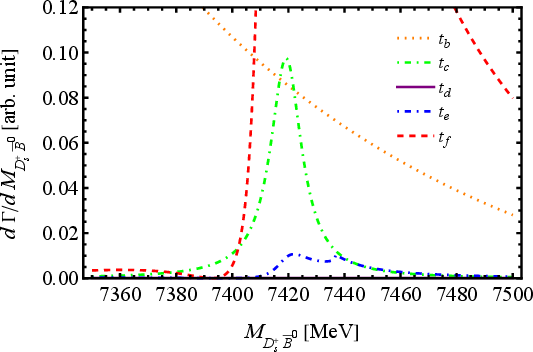}\label{fig:ls-550-b}}
    \caption{(a) The $\bar{B}^0 D_s^+$ invariant mass distribution for the process $\Upsilon \to D^{-} \bar{B}^{0} D_s^{+}$, calculated using the parameters from Scheme II. A threshold enhancement is observed near the $\bar{B}^0 D_s^+$ threshold, and a clear peak appears at approximately $7415\,\text{MeV}$, which corresponds to the contributions of the $T_{bc\bar{s}}^{0,PP}$ and $T_{bc\bar{s}}^{2,VV}$ states, respectively. (b) The region of (a) enclosed by the cyan dashed line, highlighting the relative magnitudes of the individual amplitudes $t_b, t_c, \dots, t_f$. The legend entries $t_a, t_b, \dots, t_f$ represent the partial contributions from the amplitudes $\mathcal{M}^{a}_{\text{tree}}$, $\mathcal{M}^{b}_{\text{loop}}$, \dots, $\mathcal{M}^{f}_{\text{loop}}$, respectively.}
    \label{fig:ls-550}
\end{figure}

In Scheme II, we again use Eq.~\eqref{eq:matchDR} to determine the finite part of the loop amplitude within the dimensional regularization framework. To maintain consistency with the analysis in Appendix~\ref{sec:app-a-2}, the three-momentum cutoff parameter is set to $q_{\text{max}} = 550\,\text{MeV}$ in the corresponding cutoff scheme. The pole positions and coupling constants of the $T_{bc\bar{s}}$ states are taken from Eq.~\eqref{eq:coupling-II} and Table~\ref{tab:results2} in Appendix~\ref{sec:app-a-2}.

As in Scheme I, the $\bar{B}^0 D_s^+$ invariant mass distribution exhibits a peak near $7415\,\text{MeV}$ and a dip at the $\bar{B}^{*0} D_s^{*+}$ threshold. However, the increased cutoff leads to stronger couplings, resulting in more pronounced dynamical effects. Notably, the amplitude $t_c$, associated with the rescattering process $\Upsilon \to D^{-} \bar{B}^{0} D_s^{+} \to D^{-} T_{bc\bar{s}}^{0,VV} \to D^{-} \bar{B}^{0} D_s^{+}$, now produces a visible dip structure near $7420\,\text{MeV}$. This feature arises from destructive interference between $t_c$ and the dominant tree-level amplitude $t_a$, although the magnitude ratio between the amplitudes $t_f$, $t_a$ and $t_c$ remains almost unchanged. The emergence of this structure arises from the increased difference in the pole positions between $T_{bc\bar{s}}^{0,VV}$ and $T_{bc\bar{s}}^{2,VV}$.

In addition, the amplitude $t_b$, corresponding to the process $\Upsilon \to D^{-} \bar{B}^{0} D_s^{+} \to D^{-} T_{bc\bar{s}}^{0,PP} \to D^{-} \bar{B}^{0} D_s^{+}$, exhibits a significantly larger contribution near threshold. After interference with the tree amplitude, this results in a clear threshold enhancement in the final distribution. This contrasts with Scheme I, where such an effect was absent, and underscores the sensitivity of the lineshape to the strength of the $T_{bc\bar{s}}^{0,PP}$ coupling, which increases with the larger cutoff value. 

\section{Summary}
\label{sec:concl}

In this work we have investigated the weak decay process $\Upsilon \to D^{-} \bar{B}^{0} D_s^{+}$ as a potential probe for the doubly heavy tetraquark molecular states $T_{bc\bar{s}}$. By combining Fierz rearrangement with  final-state interactions, we have derived the dominant decay amplitudes and computed the corresponding $\bar{B}^0 D_s^+$ invariant mass distribution.

Compared to our previous study \cite{Liu:2023hrz}, we have improved the pole positions of $T_{bc\bar{s}}$ states by incorporating box-diagram. Within two distinct regularization schemes, labeled Scheme I ($q_{\text{max}} = 450\,\text{MeV}$) and Scheme II ($q_{\text{max}} = 550\,\text{MeV}$), we have recalculated the pole positions and coupling strengths of the $T_{bc\bar{s}}$ states with various quantum numbers and hadronic components, as detailed in Appendix~\ref{sec:app-a}. Using these inputs, we have computed the invariant mass spectra for $\Upsilon \to D^{-} \bar{B}^{0} D_s^{+}$, with the results presented in Figs.~\ref{fig:ls-450} and \ref{fig:ls-550}, respectively.

Our analysis reveals a prominent peak structure near $7425\,\text{MeV}$ in scheme I and $7415\,\text{MeV}$ in scheme II, which is predominantly attributed to the $T_{bc\bar{s}}^{2,VV}$ state. In contrast, the contribution of the $T_{bc\bar{s}}^{0,VV}$ state, while present, is obscured due to its proximity in mass to the $T_{bc\bar{s}}^{2,VV}$ state. As a result, it does not appear as a distinct peak but may instead manifest as a subtle dip structure arising from interference with the non-resonance background.

The behavior of the scalar $T_{bc\bar{s}}^{0,PP}$ state is found to be sensitive to the choice of parameters. In Scheme II, where the coupling is enhanced, a clear near-threshold enhancement is observed, whereas no such feature appears in Scheme I. This indicates that the detectability of $T_{bc\bar{s}}^{0,PP}$ depends critically on the strength of the underlying hadron-hadron interaction and the regularization scale.


In this work we employ local currents and naive factorization scheme to study $\Upsilon$ weak decay. As a result, our theoretical uncertainties are expected to be larger compared to those derived from the well-established QCD factorization approach \cite{Beneke:1999br,Beneke:2000ry,Beneke:2001ev}. Nevertheless, our focus is not on the precise prediction of absolute decay widths, but rather on the characteristic structures in the invariant mass distribution induced by the $T_{bc\bar{s}}$ states. The appearance of these features is predominantly determined by the relative contributions among different production mechanisms, $\mathcal{M}^{a}_{\text{tree}}$, $\mathcal{M}^{b}_{\text{loop}}$, \dots, $\mathcal{M}^{f}_{\text{loop}}$. Consequently, many of the theoretical uncertainties cancel out, leading to a reliable interpretation of the observed structures.

In conclusion, the decay $\Upsilon \to D^{-} \bar{B}^{0} D_s^{+}$ provides a promising scenario for searching for doubly heavy tetraquark molecular states $T_{bc\bar{s}}$. In particular, the $T_{bc\bar{s}}^{2,VV}$ state produces a robust and identifiable peak, making it an ideal candidate for experimental discovery. Meanwhile, $T_{bc\bar{s}}^{0,PP}$ state and $T_{bc\bar{s}}^{0,VV}$ state may exhibit a threshold enhancement and a dip structure, respectively. Future measurements by experiments could exploit this channel to confirm the existence of $T_{bc\bar{s}}$ states and probe their internal structure.

\begin{acknowledgments}
    We would like to express our sincere gratitude to Prof. Eulogio Oset for his valuable suggestions and insightful comments on this paper. This project is supported by the National Natural Science Foundation of China under Grant No.~12075019, the Jiangsu Provincial Double-Innovation Program under Grant No.~JSSCRC2021488, and the Fundamental Research Funds for the Central Universities.
\end{acknowledgments}

\bibliography{ref}

%


\appendix
\section{Pole positions and coupling constants of the possible $T_{bc \bar s}$ states}
\label{sec:app-a}


In this section we refine the analysis presented in our previous work~\cite{Liu:2023hrz} by incorporating the exchange of light pseudoscalar mesons, which allows us to compute the decay widths of the candidate hadronic molecular states identified therein. In Ref.~\cite{Liu:2023hrz}, six possible $T_{bc\bar{s}}$ molecular states were found, with binding energies in the range of 10–20 $\mev$ when the cutoff parameter was set to $q_{\text{max}} = 600\,\text{MeV}$.

A similar approach was employed in Ref.~\cite{Feijoo:2021ppq} to study the $T_{cc}$ state as a molecular candidate arising from $D^*D$ interactions. To reproduce the observed binding energy of $T_{cc}$, the authors determined that the cutoff $q_{\text{max}}$  lie between 415 and 476 $\mev$, with the uncertainty largely due to the difficulty in quantifying contributions from heavy-meson exchanges.

Motivated by this analysis, we evaluate the decay widths of the $T_{bc\bar{s}}$ candidate states in two representative scenarios: (I) $q_{\text{max}} = 450\,\text{MeV}$ and (II) $q_{\text{max}} = 550\,\text{MeV}$. These values are chosen to bracket the physically motivated range suggested by analogous systems like $T_{cc}$, while also allowing us to assess the dependence of the resonance properties and decay patterns on the cutoff scale. 

\begin{figure}[htb]
    \centering
    \subfigure[]{\includegraphics[scale=0.5]{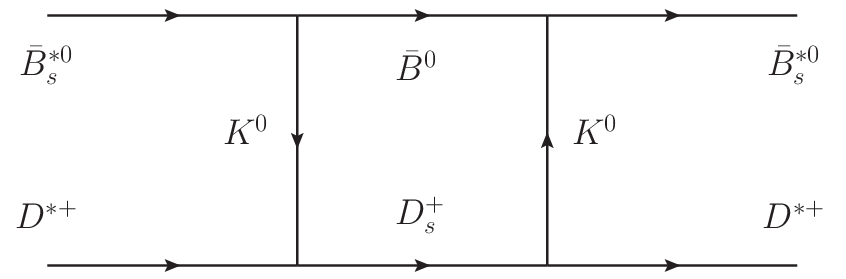}\label{fig:box-a1}}
    \subfigure[]{\includegraphics[scale=0.5]{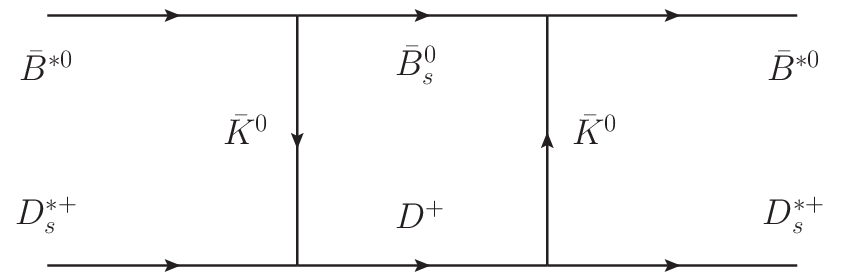}\label{fig:box-a2}} \\
    \subfigure[]{\includegraphics[scale=0.5]{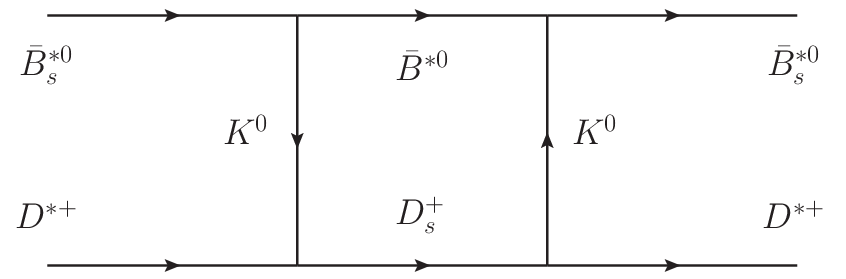}\label{fig:box-b1}}
    \subfigure[]{\includegraphics[scale=0.5]{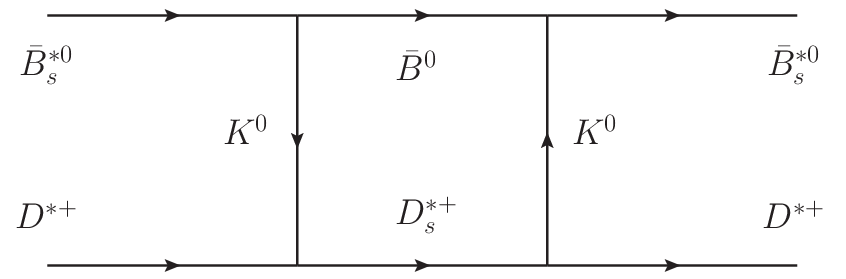}\label{fig:box-b2}} \\
    \subfigure[]{\includegraphics[scale=0.5]{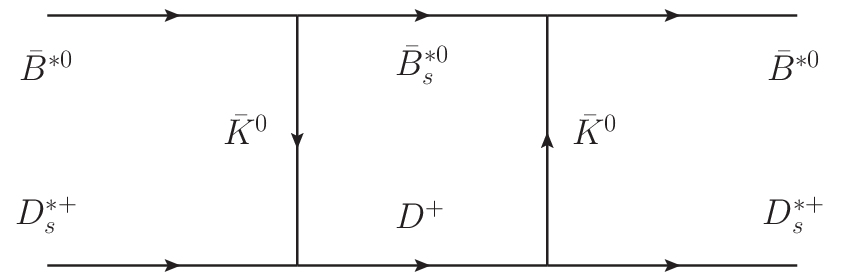}\label{fig:box-b3}}
    \subfigure[]{\includegraphics[scale=0.5]{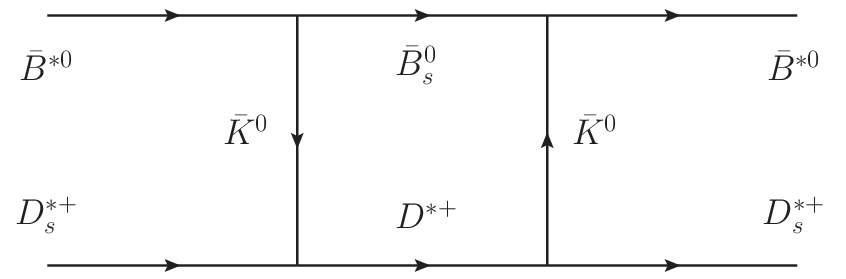}\label{fig:box-b4}}
    \caption{Box diagrams for the evaluation of the width of the $T_{bc \bar s}$ states.}
    \label{fig:box}
\end{figure}

Building upon the previous framework, we improve the analysis by incorporating box diagrams involving the exchange of kaon mesons. This extension allows the previously identified bound-state poles to acquire a finite width, thereby providing a more realistic description of the resonance properties.

Following the procedure detailed in Ref.~\cite{Oset:2022xji}, we take into account the contributions of box diagrams shown in Fig.~\ref{fig:box}. For diagram.~\ref{fig:box-a1} and ~\ref{fig:box-a2} , the box-diagram contribution to the potential is given by:
\begin{align}
    V_{\text{box}}^{a,b} = i \, 16 g^4 \int \frac{d^4 q}{(2\pi)^4} \frac{ (\vec{\epsilon}_1 \cdot \vec{q}) (\vec{\epsilon}_2 \cdot \vec{q}) (\vec{\epsilon}_3 \cdot \vec{q}) (\vec{\epsilon}_4 \cdot \vec{q}) }{(q^2 - m_K^2)^2 \left((p_1 - q)^2 - m_1^2 + i\epsilon\right) \left((p_2 + q)^2 - m_2^2 + i\epsilon\right)} \,,
    \label{eq:V-ab}
\end{align}
where $p_1$ and $p_2$ denote the four-momenta of the incoming mesons. The masses $m_1$ and $m_2$ correspond to the intermediate pseudoscalar mesons: for diagram.~\ref{fig:box-a1}, these are $m_{\bar{B}^0}$ and $m_{D_s^+}$, respectively; for diagram.~\ref{fig:box-a2}, they are $m_{\bar{B}_s^0}$ and $m_{D^+}$. The coupling constant $g$ appearing in the Eq.~(\ref{eq:V-ab}) originates from the vector-pseudoscalar-pseudoscalar $(VPP)$ vertex, which is described by the following Lagrangian derived from the Hidden Gauge Symmetry (HGS) formalism \cite{Bando:1984ej}:  
\begin{align}
  \mathcal{L}_{VPP} = -ig \, \langle [P, \partial_{\mu} P] V^{\mu} \rangle \, ,
\end{align}
where $g = \frac{M_V}{2f}$, with $M_V = 800\,\text{MeV}$ and $f = 93\,\text{MeV}$.

We evaluate the imaginary part of the box potential, which determines the decay width of the molecular states. The result is:
\begin{align}
    \text{Im}\, V_{\text{box}}^{a,b} = - \frac{4}{3} g^4 \frac{1}{10\pi \sqrt{s}} q^5 \frac{1}{\left[(\omega_2 - p_2^0)^2 - \vec{q}^2 - m_K^2\right]^2} F_J F^4(q) \left(\frac{M_1}{m_{K^*}}\right)^2 \left(\frac{M_2}{m_{K^*}}\right)^2 ,
    \label{eq:box-a-im}
\end{align}
where $\omega_2 = \sqrt{m_2^2 + \vec{q}^2}$, and $q = \lambda^{1/2}(s, m_1^2, m_2^2)/(2\sqrt{s})$ is the three-momentum of the intermediate mesons in the center-of-mass frame. The quantity $p_2^0 = (s + m_2^2 - m_1^2)/(2\sqrt{s})$ represents the energy component of the intermediate meson. The factor $F_J$ accounts for spin dependence coefficients: $F_J = 5$ for $J=0$, $F_J = 0$ for $J=1$, and $F_J = 2$ for $J=2$.

In Eq.~\eqref{eq:box-a-im}, we include a factor $M_{H^*} / m_{K^*}$ for each $H^* H K$ vertex, where $H^{(*)}$ represent the heavy flavour hadron and $M_{H^*}$ is the mass of the external heavy vector meson. This factor arises from the normalization of the vector meson fields and is discussed in detail in Ref.~\cite{Liang:2014eba}. Additionally, to account for the off-shell nature of the exchanged kaon, we introduce a form factor at each vertex, following Ref.~\cite{Molina:2010tx}, of the form $F(q) = \exp(-q^2 / \Lambda^2)$, with $\Lambda \sim 1200\,\text{MeV}$ \cite{Navarra:2001ju}. In our case, the loop momentum dependence is approximated as $F(q) = \exp\left[-\left((\omega_K - p_2^0)^2 - \vec{q}^2\right)/\Lambda^2\right]$, where $\omega_K = \sqrt{m_K^2 + \vec{q}^2}$.

The contributions to the imaginary part of the potential from the box diagrams in Fig.~\ref{fig:box}~(c), (d), (e), and (f) were originally computed in the context of $D^* \bar{K}^*$ interactions in Ref.~\cite{Molina:2020hde}. In the present case, the corresponding amplitudes can be obtained by appropriately replacing the hadron masses, leading to analogous expressions:
\begin{align}
\text{Im}\, V_{\text{box}}^{c,e} &= -\frac{4}{9} \frac{1}{8\pi\sqrt{s}} \, q^5 \left(G' g M_1 \right)^2 \frac{1}{\left[(\omega_2 - p_2^0)^2 - \vec{q}^2 - m_K^2\right]^2} F'_J F^4(q) \left(\frac{M_2}{m_{K^*}}\right)^2 \, , \nonumber \\
\text{Im}\, V_{\text{box}}^{d,f} &= -\frac{4}{9} \frac{1}{8\pi\sqrt{s}} \, q^5 \left(G' g M_2 \right)^2 \frac{1}{\left[(\omega_2 - p_2^0)^2 - \vec{q}^2 - m_K^2\right]^2} F'_J F^4(q) \left(\frac{M_1}{m_{K^*}}\right)^2 \,,
\label{eq:boxB}
\end{align}
where $F'_J = 0$, $3/2$, and $9/10$ for total angular momentum $J = 0$, $1$, and $2$, respectively. The kinematic variables are defined as $q = \lambda^{1/2}(s, m_1^2, m_2^2)/(2\sqrt{s})$ and $p_2^0 = (s + m_2^2 - m_1^2)/(2\sqrt{s})$, with $\omega_2 = \sqrt{m_2^2 + \vec{q}^2}$. As in Eq.~(\ref{eq:box-a-im}), the factor $M_{H^*} / m_{K^*}$ in Eq.~(\ref{eq:boxB}) arises from the $H^* H K$ vertex.

In Eq.~(\ref{eq:boxB}), the vector-vector-pseudoscalar ($VVP$) vertex is incorporated based on the Lagrangian proposed in Refs.~\cite{Oset:2022xji,Bramon:1992kr}, given by  
\begin{align}
  \mathcal{L}_{VVP} = \frac{G^\prime}{\sqrt{2}} \epsilon^{\mu\nu\alpha\beta} \, \langle \partial_{\mu} V_{\nu} \partial_{\alpha} V_{\beta} P \rangle \, ,
\end{align}
where the coupling constant $G^\prime$ is defined as $G^\prime = \frac{3g^{\prime 2}}{4\pi^2 f} $ with $g^\prime = -\frac{G_V M_\rho}{\sqrt{2} f^2} $, $f = 93\,\text{MeV}$ and $G_V = 55\,\text{MeV}$. It is worth noting that the heavy-quark spin symmetry factor associated with $M_{1}$ or $M_{2}$ in the coefficient $G' g M_H$ is naturally incorporated through the anomalous tensor coupling at the $B^* B^* \pi$ vertex, as discussed in Ref.~\cite{Oset:2022xji}.


The full potential $V_{\text{full}}$ entering the Bethe-Salpeter equation is then constructed by adding the calculated imaginary parts of the box diagrams to $V_{\text{contact}}$ and $V_{\text{exch.}}$ calculated before:
\begin{align}
    V_{\text{full}} = V_{\text{contact}} + V_{\text{exch.}} + i\,\text{Im}\, V_{\text{box}}^{a} + i\,\text{Im}\, V_{\text{box}}^{b} + i\,\text{Im}\, V_{\text{box}}^{c} + i\,\text{Im}\, V_{\text{box}}^{d} + i\,\text{Im}\, V_{\text{box}}^{e} + i\,\text{Im}\, V_{\text{box}}^{f} \,.
    \label{eq:Vfull}
\end{align}
Here, $V_{\text{contact}}$ and $V_{\text{exch.}}$ denote the contact and vector exchange potentials taken from the previous work \cite{Liu:2023hrz}. By replacing the interaction potential $V$ in Ref.~\cite{Liu:2023hrz} with $V_{\text{full}}$ in Eq.~(\ref{eq:Vfull}) and repeating the calculation procedure of Ref.~\cite{Liu:2023hrz}, we can obtain the improved pole positions of the $T_{bc\bar{s}}$ molecular states.

\subsection{scenario I : $q_{max} =450$ MeV}
\label{sec:app-a-1}


We follow the same procedure as in Ref.~\cite{Liu:2023hrz}, setting the three-momentum cutoff to $q_{\text{max}} = 450\,\text{MeV}$, to determine the pole positions of the $T_{bc\bar{s}}$ molecular states. The resulting values, along with the corresponding coupling strengths to the relevant hadronic channels, are summarized in Table~\ref{tab:results1}. 

\begin{table}[htb]
\centering
\renewcommand{\arraystretch}{1.5}
\caption{The pole position $E_{\text{pole}}$ and the couplings $g_i$ of the bound states on the physical (first Riemann) sheet for the $bc\bar{s}\bar{d}$ systems, with the cut-off momentum $q_\textrm{max}=450\mev$.}
\begin{tabular}{ccccc}
\hline\hline
Content: $bc\bar{s}\bar{d}$ & \,$I(J^P)$\, & $E_{\text{pole}}$ & \,Channel\, & $|g_i|$ (GeV)
\\ \hline\hline
\multirow{2}{*}{$T_{bc\bar s}^{0,PP}$} & \multirow{2}{*}{$\frac{1}{2}(0^+)$} & \multirow{2}{*}{7236.2 - i0} & $\bar{B}_s^0D^+$        & 7.9
\\ \cline{4-5}
                                                                                                              &&& $\bar{B}^0D_s^+$        & 10.4  	
\\ \hline
\multirow{2}{*}{$T_{bc\bar s}^{1,VP}$} & \multirow{2}{*}{$\frac{1}{2}(1^+)$} & \multirow{2}{*}{7284.0 - i0}  & $\bar{B}_s^{*0}D^+$  & 9.5
\\ \cline{4-5}
                                                                                                                 &&& $\bar{B}^{*0}D_s^+$  & 11.6  	
\\ \hline
\multirow{2}{*}{$T_{bc\bar s}^{1,VP^\prime}$} & \multirow{2}{*}{$\frac{1}{2}(1^+)$} & \multirow{2}{*}{7376.9 - i0}  & $\bar{B}_s^0D^{*+}$  & 7.8
\\ \cline{4-5}
                                                                                                                 &&& $\bar{B}^0D_s^{*+}$  & 10.9  	

\\ \hline
\multirow{2}{*}{$T_{bc\bar s}^{0,VV}$} & \multirow{2}{*}{$\frac{1}{2}(0^+)$} & \multirow{2}{*}{7425.5 - i2.8} & $\bar{B}_s^{*0}D^{*+}$ & 11.3
\\ \cline{4-5}
                                                                                                                  &&& $\bar B^{*0}D_s^{*+}$       & 14.2
\\ \hline
\multirow{2}{*}{$T_{bc\bar s}^{1,VV}$} & \multirow{2}{*}{$\frac{1}{2}(1^+)$} & \multirow{2}{*}{7425.1 - i1.9} & $\bar{B}_s^{*0}D^{*+}$ & 11.3
\\ \cline{4-5}
                                                                                                                  &&& $\bar B^{*0}D_s^{*+}$       & 14.1
\\ \hline
\multirow{2}{*}{$T_{bc\bar s}^{2,VV}$} & \multirow{2}{*}{$\frac{1}{2}(2^+)$} & \multirow{2}{*}{7424.5 - i2.5} & $\bar{B}_s^{*0}D^{*+}$ & 12.2
\\ \cline{4-5}
                                                                                                                  &&& $\bar B^{*0}D_s^{*+}$       & 15.0
\\ \hline\hline

\end{tabular}
\label{tab:results1}
\end{table}

By incorporating the contributions from the box diagrams—see Eqs.~\eqref{eq:box-a-im} and \eqref{eq:boxB}—the previously bound states acquire a finite width through their coupling to open decay channels. This allows us to evaluate the total decay width $\Gamma_X$ of each resonance $X$ as $\Gamma_X = 2\,|\text{Im}\,E_{\text{pole}}|$, where $E_{\text{pole}}$ is the complex energy of the pole in the scattering amplitude $T$. Furthermore, we can decompose the total width into partial widths for specific final states by analyzing the relative strength of the imaginary parts of the interaction potential:
\begin{align}
    \Gamma_{X \to \bar B^0 D_s^+} =& \Gamma_X \times {\text{Im} \, V_{\text{box}}^a (s) \over \text{Im} \, V(s) } \big |_{s=E^2_{\text{pole}}} \, , \\
    \Gamma_{X \to \bar B_s^0 D^+} =& \Gamma_X \times {\text{Im} \, V_{\text{box}}^b (s) \over \text{Im} \, V(s) } \big |_{s=E^2_{\text{pole}}} \, , \\
    \Gamma_{X \to \bar B^{*0} D_s^+} =& \Gamma_X \times {\text{Im} \, V_{\text{box}}^c (s) \over \text{Im} \, V(s) } \big |_{s=E^2_{\text{pole}}} \, , \\
    \Gamma_{X \to \bar B^0 D_s^{*+}} =& \Gamma_X \times {\text{Im} \, V_{\text{box}}^d (s) \over \text{Im} \, V(s) } \big |_{s=E^2_{\text{pole}}} \, , \\
    \Gamma_{X \to \bar B_s^{*0} D^+} =& \Gamma_X \times {\text{Im} \, V_{\text{box}}^e (s) \over \text{Im} \, V(s) } \big |_{s=E^2_{\text{pole}}} \, , \\
    \Gamma_{X \to \bar B_s^0 D^{*+}} =& \Gamma_X \times {\text{Im} \, V_{\text{box}}^f (s) \over \text{Im} \, V(s) } \big |_{s=E^2_{\text{pole}}} \, .
\end{align}

Here, we present the general coupling Lagrangian between the molecular states and their decay channels, including $ X_{VV}^0 $ coupled to two pseudoscalar mesons (denoted as $ \mathcal{L}_{PPX_{VV}^0} $), $ X_{VV}^1 $ coupled to one pseudoscalar and one vector meson (denoted as $ \mathcal{L}_{P V_\mu X_{VV}^{1,\mu}} $), and $ X_{VV}^2 $ coupled to two pseudoscalar mesons (denoted as $ \mathcal{L}_{ P  P  X_{VV}^{2,\mu\nu}} $). In the equations, $X_{VV}^0$ denotes $T_{bc \bar s}^{0,VV}$, the molecular state with angular momentum $J = 0$ dynamically generated from $\bar{B}_s^{*0} D^{*+} \,\text{-}\, \bar{B}^{*0} D_s^{*+}$  vector-vector meson interactions. Similar shorthand notations will be used consistently throughout this section.
\begin{align}
    \mathcal{L}_{PPX_{VV}^0} =& g \, P P X_{VV}^0 \, ,
    \label{eq:L-coup-PPX0} \\
    \mathcal{L}_{P V_\mu X_{VV}^{1,\mu}} =& g \, P V_\mu X_{VV}^{1,\mu} \, ,
    \label{eq:L-coup-PPX1} \\
    \mathcal{L}_{ P  P  X_{VV}^{2,\mu\nu}} =& g \, \partial_\mu P \partial_\nu P X_{VV}^{2,\mu\nu} \, .
    \label{eq:L-coup-PPX2}
\end{align}



Using these Lagrangians, we derive the partial decay widths in terms of the coupling constant $g$ and phase space. For the $0^+$ state:
\begin{equation}
    \Gamma_{X_{VV}^0 \to \bar{B}^0 D_s^+ / \bar{B}_s^0 D^+} = \frac{1}{8\pi} \frac{g^2}{m_X^2} \, k \,,
\end{equation}
and for the $2^+$ state:
\begin{equation}
    \Gamma_{X_{VV}^2 \to \bar{B}^0 D_s^+ / \bar{B}_s^0 D^+} = \frac{2}{15} \cdot \frac{1}{8\pi} \frac{g^2}{m_X^2} \, k^5 \,,
\end{equation}
where $k = \sqrt{(m_X^2 - (m_1 + m_2)^2)(m_X^2 - (m_1 - m_2)^2)}/(2m_X)$ is the center-of-mass momentum of the outgoing mesons.

Using the above relations, we extract the partial widths and coupling constants as:
\begin{align}
    \Gamma_{X_{VV}^0 \to \bar{B}^0 D_s^+} &= 2.71\,\text{MeV}, \quad\quad\quad \Gamma_{X_{VV}^0 \to \bar{B}_s^0 D^+} = 2.89\,\text{MeV}, \nonumber \\
    \Gamma_{X_{VV}^2 \to \bar{B}^0 D_s^+} &= 1.15\,\text{MeV}, \quad\quad\quad \Gamma_{X_{VV}^2 \to \bar{B}_s^0 D^+} = 1.23\,\text{MeV}, \\
    g_{X_{VV}^0 \to \bar{B}^0 D_s^+} &= 2281\,\text{MeV}, \quad\quad\;\;\, g_{X_{VV}^0 \to \bar{B}_s^0 D^+} = 2263\,\text{MeV}, \nonumber \\
    g_{X_{VV}^2 \to \bar{B}^0 D_s^+} &= 7.86 \times 10^{-3}\,\text{MeV}^{-1}, \quad g_{X_{VV}^2 \to \bar{B}_s^0 D^+} = 7.55 \times 10^{-3}\,\text{MeV}^{-1}.
    \label{eq:coupling-I}
\end{align}


\subsection{scenario II : $q_{max} =550$ MeV}
\label{sec:app-a-2}

\begin{table}[htb]
\centering
\renewcommand{\arraystretch}{1.5}
\caption{The pole position $E_{\text{pole}}$ and the couplings $g_i$ of the bound states on the physical (first Riemann) sheet for the $bc\bar{s}\bar{d}$ systems, with the cut-off momentum $q_\textrm{max}=550\mev$.}
\begin{tabular}{ccccc}
\hline\hline
Content: $bc\bar{s}\bar{d}$ & \,$I(J^P)$\, & $E_{\text{pole}}$ & \,Channel\, & $|g_i|$ (GeV)
\\ \hline\hline
\multirow{2}{*}{$T_{bc\bar s}^{0,PP}$} & \multirow{2}{*}{$\frac{1}{2}(0^+)$} & \multirow{2}{*}{7228.1 - i0} & $\bar{B}_s^0D^+$        & 16.1
\\ \cline{4-5}
                                                                                                              &&& $\bar{B}^0D_s^+$        & 18.3  	
\\ \hline
\multirow{2}{*}{$T_{bc\bar s}^{1,VP}$} & \multirow{2}{*}{$\frac{1}{2}(1^+)$} & \multirow{2}{*}{7275.1 - i0}  & $\bar{B}_s^{*0}D^+$  & 16.8
\\ \cline{4-5}
                                                                                                                 &&& $\bar{B}^{*0}D_s^+$  & 18.4  	
\\ \hline
\multirow{2}{*}{$T_{bc\bar s}^{1,VP^\prime}$} & \multirow{2}{*}{$\frac{1}{2}(1^+)$} & \multirow{2}{*}{7368.4 - i0}  & $\bar{B}_s^0D^{*+}$  & 16.8
\\ \cline{4-5}
                                                                                                                 &&& $\bar{B}^0D_s^{*+}$  & 19.6  	

\\ \hline
\multirow{2}{*}{$T_{bc\bar s}^{0,VV}$} & \multirow{2}{*}{$\frac{1}{2}(0^+)$} & \multirow{2}{*}{7419.2 - i6.8} & $\bar{B}_s^{*0}D^{*+}$ & 16.5
\\ \cline{4-5}
                                                                                                                  &&& $\bar B^{*0}D_s^{*+}$       & 18.6
\\ \hline
\multirow{2}{*}{$T_{bc\bar s}^{1,VV}$} & \multirow{2}{*}{$\frac{1}{2}(1^+)$} & \multirow{2}{*}{7415.2 - i4.7} & $\bar{B}_s^{*0}D^{*+}$ & 18.0
\\ \cline{4-5}
                                                                                                                  &&& $\bar B^{*0}D_s^{*+}$       & 20.2
\\ \hline
\multirow{2}{*}{$T_{bc\bar s}^{2,VV}$} & \multirow{2}{*}{$\frac{1}{2}(2^+)$} & \multirow{2}{*}{7413.7 - i5.8} & $\bar{B}_s^{*0}D^{*+}$ & 19.0
\\ \cline{4-5}
                                                                                                                  &&& $\bar B^{*0}D_s^{*+}$       & 21.1
\\ \hline\hline

\end{tabular}
\label{tab:results2}
\end{table}



Following the same procedure as in Ref.~\cite{Liu:2023hrz}, we perform the calculation with a three-momentum cutoff of $q_{\text{max}} = 550\,\text{MeV}$. This yields the pole positions of the $T_{bc\bar{s}}$ molecular states and the corresponding coupling constants to the coupled meson-meson channels. The results are summarized in Table~\ref{tab:results2}.

Using the imaginary parts of the box-diagram contributions and the effective Lagrangians introduced in Sec.~\ref{sec:app-a-1}, we extract the partial decay widths and hadronic coupling constants. The results are:
\begin{align}
    \Gamma_{X_{VV}^0 \to \bar{B}^0 D_s^+} &= 6.58\,\text{MeV}, \quad\quad\quad\quad \Gamma_{X_{VV}^0 \to \bar{B}_s^0 D^+} = 7.02\,\text{MeV}, \nonumber \\
    \Gamma_{X_{VV}^2 \to \bar{B}^0 D_s^+} &= 2.86\,\text{MeV}, \quad\quad\quad\quad \Gamma_{X_{VV}^2 \to \bar{B}_s^0 D^+} = 3.06\,\text{MeV}, \\
    g_{X_{VV}^0 \to \bar{B}^0 D_s^+} &= 3584\,\text{MeV}, \quad\quad\quad\;\;\, g_{X_{VV}^0 \to \bar{B}_s^0 D^+} = 3553\,\text{MeV}, \nonumber \\
    g_{X_{VV}^2 \to \bar{B}^0 D_s^+} &= 1.34 \times 10^{-2}\,\text{MeV}^{-1}, \quad g_{X_{VV}^2 \to \bar{B}_s^0 D^+} = 1.32 \times 10^{-2}\,\text{MeV}^{-1},
    \label{eq:coupling-II}
\end{align}
where $X_{VV}^0 \equiv T_{bc\bar{s}}^{0,VV}$ and $X_{VV}^2 \equiv T_{bc\bar{s}}^{2,VV}$. 

As expected, increasing the cutoff enhances the attraction in the coupled-channel system, leading to deeper binding and larger coupling strengths compared to Scenario I ($q_{\text{max}} = 450\,\text{MeV}$). This results in broader decay widths, reflecting a stronger coupling to the open channels. These parameters are used in the main analysis to assess the sensitivity of the invariant mass distribution to the underlying hadron-hadron interaction strength.

\end{document}